\documentclass[11pt,letterpaper]{article}
\usepackage{naaclhlt2015}
\usepackage{mystuff-nlp,color}
\usepackage{times}
\usepackage{standalone}
\usepackage{url,latexsym}
\usepackage{amsmath,algpseudocode,graphicx,amssymb}

\newcommand{\vb}[0]{\vec{\beta}}
\newcommand{\ve}[0]{\vec{\eta}}

\newcommand{\Z}[0]{Z(\ve, \vb; G)}

\newcommand{\example}[1]{\emph{#1}}

\renewcommand{\jemph}[1]{\textbf{#1}}
\newcommand{\vthr}{\vth^\rightarrow}
\newcommand{\vthl}{\vth^\leftarrow}

\title{``You're Mr. Lebowski, I'm the Dude'':\\Inducing Address Term Formality in Signed Social Networks}

\author{Vinodh Krishnan\\
College of Computing\\
Georgia Institute of Technology\\
Atlanta, GA 30308\\
	    {\tt krishnan.vinodh@gmail.com}
	  \And
    Jacob Eisenstein\\
School of Interactive Computing\\
Georgia Institute of Technology\\
Atlanta, GA 30308\\
  {\tt jacobe@gatech.edu}}
\date{}

\begin{document}
\maketitle
\begin{abstract}
We present an unsupervised model for inducing signed social networks from the content exchanged across network edges. Inference in this model solves three problems simultaneously: (1) identifying the sign of each edge; (2) characterizing the distribution over content for each edge type; (3) estimating weights for triadic features that map to theoretical models such as structural balance. We apply this model to the problem of inducing the social function of \jemph{address terms}, such as \example{Madame}, \example{comrade}, and \example{dude}. On a dataset of movie scripts, our system obtains a coherent clustering of address terms, while at the same time making intuitively plausible judgments of the formality of social relations in each film. As an additional contribution, we provide a bootstrapping technique for identifying and tagging address terms in dialogue. 

\end{abstract}

\section{Introduction}
One of the core communicative functions of language is to modulate and reproduce \jemph{social dynamics}, such as friendship, familiarity, formality, and power~\cite{hymes1972communicative}. However, large-scale empirical work on understanding this communicative function has been stymied by a lack of labeled data: it is not clear what to annotate, let alone whether and how such annotations can be produced reliably. Computational linguistics has made great progress in modeling language's informational dimension, but --- with a few notable exceptions --- computation has had little to contribute to our understanding of language's social dimension.

Yet there is a rich theoretical literature on social structures and dynamics. In this paper, we focus on one such structure: signed social networks, in which edges between individuals are annotated with information about the nature of the relationship. For example, the individuals in a dyad may be friends or foes; they may be on formal or informal terms; or they may be in an asymmetric power relationship. Several theories characterize signed social networks: in structural balance theory, edge signs indicate friendship and enmity, with some triads of signed edges being stable, and others being unstable~\cite{cartwright1956structural}; conversely, in status theory~\cite{leskovec2010signed}, edges indicate status differentials, and triads should obey transitivity. But these theoretical models can only be applied when the sign of each social network connection is known, and they do not answer the sociolinguistic question of how the sign of a social tie relates to the language that is exchanged across it.

We present a unified statistical model that incorporates both network structure and linguistic content. The model connects signed social networks with \jemph{address terms}~\cite{brown1961address}, which include names, titles, and ``placeholder names,'' such as \example{dude}. The choice of address terms is an indicator of the level of formality between the two parties: for example, in contemporary North American English, a formal relationship is signaled by the use of titles such as \example{Ms} and \example{Mr}, while an informal relationship is signaled by the use of first names and placeholder names. These tendencies can be captured with a multinomial distribution over address terms, conditioned on the nature of the relationship. However, the linguistic signal is not the only indicator of formality: network structural properties can also come into play. For example, if two individuals share a mutual friend, with which both are on informal terms, then they too are more likely to have an informal relationship. With a log-linear prior distribution over network structures, it is possible to incorporate such triadic features, which relate to structural balance and status theory. 

Given a dataset of unlabeled network structures and linguistic content, inference in this model simultaneously induces three quantities of interest:

\begin{itemize}
\setlength{\itemsep}{0pt}
\item a clustering of network edges into types;
\item a probabilistic model of the address terms that are used across each edge type, thus revealing the social meaning of these address terms;
\item weights for triadic features of signed networks, which can then be compared with the predictions of existing social theories.
\end{itemize}

Such inferences can be viewed as a form of \jemph{sociolinguistic structure induction}, permitting social meanings to be drawn from linguistic data. In addition to the model and the associated inference procedure, we also present an approach for inducing a lexicon of address terms, and for tagging them in dialogues. We apply this procedure to a dataset of movie scripts~\cite{danescu2011chameleons}. Quantitative evaluation against human ratings shows that the induced clusters of address terms correspond to intuitive perceptions of formality, and that the network structural features improve predictive likelihood over a purely text-based model. Qualitative evaluation shows that the model makes reasonable predictions of the level of formality of social network ties in well-known movies.

We first describe our model for linking network structure and linguistic content in general terms, as it can be used for many types of linguistic content and edge labels. Next we describe a procedure which semi-automatically induces a lexicon of address terms, and then automatically labels them in text. We then describe the application of this procedure to a dataset of movie dialogues, including quantitative and qualitative evaluations.

\section{Joint model of signed social networks and textual content}
We now present a probabilistic model for linking network structure with content exchanged over the network. In this section, the model is presented in general terms, so that it can be applied to any type of event counts, with any form of discrete edge labels. The application of the model to forms of address is described in Sections~\ref{sec:identifying-address} and \ref{sec:address-formality}.

We observe a dataset of undirected graphs $G^{(t)} = \{ i,j \}$, with a total ordering on nodes such that $i < j$ in all edges. For each edge $\tuple{i,j}$, we observe directed content vectors $\vx_{i\to j}$ and $\vx_{i \from j}$, which may represent counts of words or other discrete events, such as up-votes and down-votes for comments in a forum thread. We hypothesize a latent edge label $y_{ij} \in \set{Y}$, so that $\vx_{i\to j}$ and $\vx_{i\from j}$ are conditioned on $y_{ij}$. In this paper we focus on binary labels (e.g., $\set{Y} = \{+,-\}$), but the approach generalizes to larger finite discrete sets, such as directed binary labels (e.g., $\set{Y} = \{++,+-,-+,--\}$) and comparative status labels (e.g., $\set{Y} = \{<, >, \approx\}$).

We model the likelihood of the observations conditioned on the edge labels as multinomial,
\begin{align}
\vx_{i\to j} \mid y_{ij} \sim & \text{Multinomial}(\vthr_{y_{ij}}) \\
\vx_{i\from j} \mid y_{ij} \sim & 
\text{Multinomial}(\vthl_{y_{ij}}).
\end{align}

Parameter tying can be employed to handle special cases. For example, if the edge labels are undirected, then we add the constraint $\vthr_y = \vthl_y, \forall y$. If the edge labels reflect relative status, then we would instead add the constraints $(\vthr_{<} = \vthl_{>})$, $(\vthr_{>} = \vthl_{<})$, and $(\vthr_{\approx} = \vthl_{\approx})$. 

The distribution over edge labelings $P(\vy)$ is modeled in a log-linear framework, with features that can consider network structure and signed triads:
\begin{align}
\notag
P(\vec{y}; G, \ve, \vb) = & \frac{1}{\Z} \\
\notag
& \times \exp \sum_{\tuple{i,j} \in G} \trans{\ve}\vf(y_{ij},i,j,G) \\
& \times \exp \sum_{\tuple{i,j,k} \in \set{T}(G)} \beta_{y_{ij},y_{jk},y_{ik}},
\label{eq:prior}
\end{align}
where $\set{T}(G)$ is the set of triads in the graph $G$. The first
term of Equation~\ref{eq:prior} represents a normalizing constant. The second term includes weights $\ve$, which apply to network features $\vf(y_{ij},i,j,G)$. This can include features like the number of mutual friends between nodes $i$ and $j$, or any number of more elaborate structural features~\cite{liben2007link}. For example, the feature weights $\ve$ could ensure that the edge label $Y_{ij} = +$ is especially likely when nodes $i$ and $j$ have many mutual friends in $G$. However, these features cannot consider any edge labels besides $y_{ij}$.

In the third line of Equation~\ref{eq:prior}, each weight $\beta_{y_{ij},y_{jk},y_{ik}}$ corresponds to a signed triad type, invariant to rotation. In a binary signed network, structural balance theory would suggest positive weights for $\beta_{+++}$ (all friends) and $\beta_{+--}$ (two friends and a mutual enemy), and negative weights for $\beta_{++-}$ (two enemies and a mutual friend) and $\beta_{---}$ (all enemies). In contrast, a status-based network theory would penalize non-transitive triads such as $\beta_{>><}$. Thus, in an unsupervised model, we can examine the weights to learn about the semantics of the induced edge types, and to see which theory best describes the signed network configurations that follow from the linguistic signal. This is a natural next step from prior work that computes the frequency of triads in explicitly-labeled signed social networks~\cite{leskovec2010signed}.

\section{Inference and estimation}
Our goal is to estimate the parameters $\theta$, $\beta$, and $\vec{\eta}$, given observations of network structures $G^{(t)}$ and linguistic content $\vx^{(t)}$, for $t \in \{1,\ldots,T\}$. Eliding the sum over instances $t$, we seek to maximize the lower bound on the expected likelihood,
\begin{align}
  \notag
  \mathcal{L}_Q = & E_Q[\log P(\vy, \vx; \vb, \vth, G)] - E_Q[\log Q(\vec{y})]\\
  \notag
  = & E_Q[\log P(\vx \mid \vy; \vth)] + E_Q[\log P(\vy ; G, \vb, \vec{\ve})]\\
& - E_Q[\log Q(\vec{y})].
  \label{eq:objective}
\end{align}

The first and third terms factor across edges,
\begin{small}
\begin{align*}
E_Q[\log P(\vx \mid \vy; \vth)] = \sum_{\tuple{i,j} \in G} \sum_{y' \in \set{Y}} & q_{ij}(y') \trans{\vx}_{i\to j} \log \vthr_{y'}\\
+ & q_{ij}(y') \trans{\vx}_{i\from j} \log \vthl_{y'}\\
E_Q[\log Q(\vy)] = \sum_{\tuple{i,j} \in G} \sum_{y' \in \set{Y}} & q_{ij}(y') \log q(y').
\end{align*}
\end{small}

The expected log-prior $E_Q[\log P(\vy)]$ is computed from the prior distribution defined in Equation~\ref{eq:prior}, and therefore involves triads of edge labels,
\begin{small}
\begin{align*}
E_Q& [\log P(\vy; \vec{\eta}, \vec{\beta})] =  - \log Z(\ve,\vb; G) \\
& + \sum_{\tuple{i,j} \in G}
\sum_{y'}q_{ij}(y') \trans{\vec{\eta}} \vf(y',i,j,G) \\
& +  \sum_{\tuple{i,j,k} \in \set{T}(G)} \sum_{y,y',y''} q_{ij}(y)q_{jk}(y') q_{ik}(y'') \beta_{y,y',y''}.
\end{align*}
\end{small}

We can reach a local maximum of the variational bound by applying expectation-maximization~\cite{dempster1977maximum}, iterating between updates to $Q(\vy)$, and updates to the parameters $\vth,\vb,\ve$. This procedure is summarized in Table~\ref{tab:em-alg}, and described in more detail below.

\subsection{E-step}
In the E-step, we sequentially update each $q_{ij}$, taking the derivative of Equation~\ref{eq:objective}:
\begin{align}
\notag
\dd{\mathcal{L}_Q}{q_{ij}(y)} = & \log P(\vx_{i\to j} \mid Y_{ij} = y; \vthr)\\
\notag
& + \log P(\vx_{i \from j} \mid Y_{ij} = y; \vthl)\\
\notag
& + E_q[\log P(\vy \mid Y_{ij} = y; \vb, \ve)]\\
& - \log q_{ij}(y) - 1.
\label{eq:dl-dq}
\end{align}

After adding a Lagrange multiplier to ensure that $\sum_y q_{ij}(y) = 1$, we obtain a closed-form solution for each $q_{ij}(y)$. These iterative updates to $q_{ij}$ can be viewed as a form of mean field inference~\cite{wainwright2008graphical}.

\subsection{M-step}
In the general case, the maximum expected likelihood solution for the content parameter $\vth$ is given by the expected counts,
\begin{align}
\label{eq:vth-r}
\vthr_{y} \propto & \sum_{\tuple{i,j} \in G}q_{ij}(y) \vx_{i\to j} \\
\label{eq:vth-l}
\vthl_{y} \propto & \sum_{\tuple{i,j} \in G}q_{ij}(y) \vx_{i\from j}.
\end{align}
As noted above, we are often interested in special cases that require parameter tying, such as $\vthr_y = \vthl_y, \forall y$. This can be handled by simply computing expected counts across the tied parameters.

\begin{table}
\hrule
\begin{enumerate}
\setlength{\itemsep}{-1pt}
\item Initialize $Q(Y^{(t)})$ for each $t \in \{1 \ldots T\}$
\item Iterate until convergence:
\begin{description}
\setlength{\itemsep}{-1pt}
\item[E-step] update each $q_{ij}$ in closed form, based on Equation~\ref{eq:dl-dq}.
\item[M-step: content] Update $\vth$ in closed form from Equations~\ref{eq:vth-r} and~\ref{eq:vth-l}.
\item[M-step: structure] Update $\vb, \ve$, and $c$ by applying L-BFGS to the noise-contrastive estimation objective in Equation~\ref{eq:nce}.
\end{description}
\end{enumerate}
\hrule
  \caption{Expectation-maximization estimation procedure}
  \label{tab:em-alg}
\end{table}

Obtaining estimates for $\vb$ and $\ve$ is more challenging, as it would seem to involve computing the partition function $\Z$, which sums over all possible labeling of each network $G^{(t)}$. The number of such labelings is exponential in the number of edges in the network. \newcite{west2014exploiting} show that for an objective function involving features on triads and dyads, it is NP-hard to find even the single optimal labeling. 

We therefore apply noise-contrastive estimation (NCE; Gutmann and Hyv\"arinen, 2012)\nocite{gutmann2012noise}, which transforms the problem of estimating the density $P(\vy)$ into a classification problem: distinguishing the observed graph labelings $\vy^{(t)}$ from randomly-generated ``noise'' labelings $\tilde{\vy}^{(t)} \sim P_n$, where $P_n$ is a noise distribution. NCE introduces an additional parameter $c$ for the partition function, so that $\log P(\vy; \vb, \ve, c) = \log P^0(\vy; \vb, \ve) + c$, with $P^0(\vy)$ representing the unnormalized probability of $\vy$. We can then obtain the NCE objective by writing $D = 1$ for the case that $\vy$ is drawn from the data distribution and $D = 0$ for the case that $\vy$ is drawn from the noise distribution, 
\begin{align}
\notag
J_{NCE}&(\ve, \vb, c)\\
\notag
= & \sum_t \log P(D = 1 \mid \vy^{(t)} ; \ve, \vb, c)\\
& - \log P(D = 0 \mid \tilde{\vy}^{(t)}; \ve, \vb, c),
\label{eq:nce}
\end{align}
where we draw exactly one noise instance $\tilde{\vy}$ for each true labeling $\vy^{(t)}$.

Because we are working in an unsupervised setting, we do not observe $\vy^{(t)}$, so we cannot directly compute the log probability in Equation~\ref{eq:nce}. Instead, we compute the expectations of the relevant log probabilities, under the distribution $Q(\vy)$,
\begin{small}
\begin{align}
\notag
E_Q&[\log P^0 (\vy; \vb, \ve)] = \\
\notag
& \sum_{\tuple{i,j} \in G} \sum_{y} q_{ij}(y) \trans{\ve}\vf(y,i,j,G) \\
& + \sum_{k : \tuple{i,j,k}\in\set{T}(G)}\sum_{y,y',y''}
q_{ij}(y) q_{jk}(y') q_{ik}(y'') \beta_{y,y',y''}.
\end{align}
\end{small}
We define the noise distribution $P_n$ by sampling edge labels $y_{ij}$ from their empirical distribution under $Q(\vec{y})$. The expectation $E_q[\log P_n(\vy)]$ is therefore simply the negative entropy of this empirical distribution, multiplied by the number of edges in $G$. We then plug in these expected log-probabilities to the noise-contrastive estimation objective function, and take derivatives with respect to the parameters $\vb$, $\ve$, and $c$. In each iteration of the M-step, we optimize these parameters using L-BFGS~\cite{liu1989limited}.

\begin{figure*}
  \centering

\begin{tabular}{lllllllllllll}
\textbf{Text}: & I & 'm & not & Mr. & Lebowski & ; & you & 're & Mr. & Lebowski & . \\
\textbf{POS}: & PRP & VBP & RB & NNP & NNP & : & PRP & VBP & NNP & NNP & .\\
\textbf{Address}: & O & O & O & B-ADDR & L-ADDR & O & O & O & B-ADDR & L-ADDR & O \\
\end{tabular}

  \caption{Automatic re-annotation of dialogue data for address term sequences}
  \label{fig:address-example}
\end{figure*}

\section{Identifying address terms in dialogue}
\label{sec:identifying-address}
The model described in the previous sections is applied in a study of the social meaning of \textbf{address terms} --- terms for addressing individual people --- which include:
\begin{description}
\setlength{\itemsep}{0pt}
\item[Names] such as \example{Barack}, \example{Barack Hussein Obama}.
\item[Titles] such as \example{Ms., Dr., Private, Reverend}. Titles can be used for address either by preceding a name (e.g., \example{Colonel Kurtz}), or in isolation (e.g., \example{Yes, Colonel}.).
\item[Placeholder names] such as \example{dude}~\cite{kiesling2004dude}, \example{bro}, \example{brother}, \example{sweetie}, \example{cousin}, and \example{asshole}. These terms can be used for address only in isolation (for example, in the address \example{cousin Sue}, the term \example{cousin} would be considered a title).
\end{description}

Because address terms connote varying levels of formality and familiarity, they play a critical role in establishing and maintaining social relationships. However, we find no prior work on automatically identifying address terms in dialogue transcripts. There are several subtasks: (1) distinguishing addresses from mentions of other individuals, (2) identifying a lexicon of titles, which either precede name addresses or can be used in isolation, (3) identifying a lexicon of placeholder names, which can only be used in isolation. We now present a tagging-based approach for performing each of these subtasks.

\begin{table}
  \small
  \centering
  \begin{tabular}{>{\bfseries}p{.3\columnwidth}p{.6\columnwidth}}
    \toprule
    Feature & Description \\
    \midrule
    Lexical & The word to be tagged, and its two predecessors and successors, $w_{i-2:i+2}$.
    \\
    POS & The part-of-speech of the token to be tagged, and the POS tags of its two predecessors and successors.\\
    Case & The case (lower, upper, or title) of the word to be tagged, and its two predessors and successors. \\
    Constituency parse & First non-NNP ancestor node of the word $w_i$ in the constituent parse tree, and all leaf node siblings in the tree.\\
    Dependency parse & All dependency relations involving $w_i$.\\
    Location & Distance of $w_i$ from the start and the end of the sentence or turn.\\
    Punctuation & All punctuation symbols occurring before and after $w_i$.\\
    Second person pronoun & All forms of the second person pronoun within the sentence.\\
    \bottomrule
  \end{tabular}
  \caption{Features used to identify address spans}
  \label{tab:address-features}
\end{table}

We build an automatically-labeled dataset from the corpus of movie dialogues provided by \newcite{danescu2011chameleons}; see Section~\ref{sec:movies} for more details. This dataset gives the identity of the speaker and addressee of each line of dialogue. These identities constitute a minimal form of manual annotation, but in many settings, such as social media dialogues, they could be obtained automatically. We augment this data by obtaining the first (given) and last (family) names of each character, which we mine from the website \url{rottentomatoes.com}. Next, we apply the CoreNLP part-of-speech tagger~\cite{manning2014stanford} to identify sequences of the NNP tag, which indicates a proper noun in the Penn Treebank Tagset~\cite{marcus1993building}. For each NNP tag sequence that contains the name of the addressee, we label it as an address, using BILOU notation~\cite{ratinov2009design}: \textbf{B}eginning, \textbf{I}nside, and \textbf{L}ast term of address segments; \textbf{O}utside and \textbf{U}nit-length sequences. An example of this tagging scheme is shown in Figure~\ref{fig:address-example}.

Next, we train a classifier (Support Vector Machine with a linear kernel) on this automatically labeled data, using the features shown in Table~\ref{tab:address-features}. For simplicity, we do not perform structured prediction, which might offer further improvements in accuracy. This classifier provides an initial, partial solution to the first problem, distinguishing second-person addresses from references to other individuals (for name references only). 
On heldout data, the classifier's macro-averaged F-measure is 83\%, and its micro-averaged F-measure is 98.7\%. Class-by-class breakdowns are shown in Table~\ref{tab:fmeasure}.

\subsection{Address term lexicons}
To our surprise, we were unable to find manually-labeled lexicons for either titles or placeholder names. We therefore employ a semi-automated approach to construct address term lexicons, bootstrapping from the address term tagger to build candidate lists, which we then manually filter.

\paragraph{Titles}
To induce a lexicon of titles, we consider terms that are frequently labeled with the tag B-ADDR across a variety of dialogues, performing a binomial test to obtain a list of terms whose frequency of being labeled as B-ADDR is significantly higher than chance. 
Of these 34 candidate terms, we manually filter out 17, which are mainly common first names, such as \example{John}; such names are frequently labeled as B-ADDR across movies. After this manual filtering, we obtain the following titles: \example{agent, aunt, captain, colonel, commander, cousin, deputy, detective, dr, herr, inspector, judge, lord, master, mayor, miss, mister, miz, monsieur, mr, mrs, ms, professor, queen, reverend, sergeant, uncle.}

\paragraph{Placeholder names} 
To induce a lexicon of placeholder names, we remove the \textsc{current-word} feature from the model, and re-run the tagger on all dialogue data. We then focus on terms which are frequently labeled \annot{U-ADDR}, indicating that they are the sole token in the address (e.g., \example{I'm}/\annot{O} \example{perfectly}/\annot{O} \example{calm}/\annot{O}, \example{dude}/\annot{U-ADDR}.) We again perform a binomial test to obtain a list of terms whose frequency of being labeled U-ADDR is significantly higher than chance. We manually filter out 41 terms from a list of 96 possible placeholder terms obtained in the previous step. Most terms eliminated were plural forms of placeholder names, such as \example{fellas} and \example{dudes}; these are indeed address terms, but because they are plural, they cannot refer to a single individual, as required by our model. Other false positives were fillers, such as \example{uh} and \example{um}, which were ocassionally labeled as \annot{I-ADDR} by our tagger. After manual filtering, we obtain the following placeholder names: \example{asshole, babe, baby, boss, boy, bro, bud, buddy, cocksucker, convict, cousin, cowboy, cunt, dad, darling, dear, detective, doll, dude, dummy, father, fella, gal, ho, hon, honey, kid, lad, lady, lover, ma, madam, madame, man, mate, mister, mon, moron, motherfucker, pal, papa, partner, peanut, pet, pilgrim, pop, president, punk, shithead, sir, sire, son, sonny, sport, sucker, sugar, sweetheart, sweetie, tiger.}

\subsection{Address term tokens}
When constructing the content vectors $\vx_{i\to j}$ and $\vx_{i\from j}$, we run the address span tagger described above, and include counts for the following types of address spans:
\begin{itemize}
\setlength{\itemsep}{0pt}
\item the bare first name, last name, and complete name of individual $j$;
\item any element in the title lexicon if labeled as B-ADDR by the tagger;
\item any element in the title or placeholder lexicon, if labeled as U-ADDR by the tagger.
\end{itemize}

\begin{table}
  \centering
  \begin{tabular}{llll}
    \toprule
    \parbox{2cm}{Class} & \parbox{2cm}{F-measure} &   Total Instances \\
    \midrule
   	\annot{I-ADDR} & 0.58 & 53\\
   	\annot{B-ADDR} & 0.800 & 483\\
   	\annot{U-ADDR} & 0.987 & 1864\\
   	\annot{L-ADDR} & 0.813 & 535\\
   	\annot{O-ADDR} & 0.993 & 35975\\
    \bottomrule
  \end{tabular}
  \caption{Breakdown of f-measure and number of instances by class in the test set.}
  \label{tab:fmeasure}
\end{table}

\section{Address terms in a model of formality}
\label{sec:address-formality}
Address terms play a key role in setting the formality of a social interaction. However, understanding this role is challenging. While some address terms, like \example{Ms} and \example{Sir}, are frequent, there is a long tail of rare terms whose meaning is more difficult to ascertain from data, such as \example{admiral}, \example{dude}, and \example{player}. Moreover, the precise social meaning of address terms can be context-dependent: for example, the term \example{comrade} may be formal in some contexts, but jokingly informal in others. 

\newcommand{\tu}[0]{\textsc{t}}
\newcommand{\vos}[0]{\textsc{v}}
Both problems can be ameliorated by adding social network structure. We treat $Y=\vos$ as indicating formality and $Y=\tu$ as indicating informality. (The notation invokes the concept of T/V systems from politeness theory~\cite{brown1987politeness}, where $\tu$ refers to the informal Latin second-person pronoun \example{tu}, and $\vos$ refers to the formal second-person pronoun \example{vos}.) 

While formality relations are clearly asymmetric in many settings, for simplicity we assume symmetric relations: each pair of individuals is either on formal or informal terms with each other. We therefore add the constraints that $\vthl_{\vos} = \vthr_{\vos}$ and $\vthl_{\tu} = \vthr_{\tu}$. In this model, we have a soft expectation that triads will obey transitivity: for example, if $i$ and $j$ have an informal relationship, and $j$ and $k$ have an informal relationship, then $i$ and $k$ are more likely to have an informal relationship. After rotation, there are four possible triads, $\annot{ttt}$, $\annot{ttv}$, $\annot{tvv}$, and $\annot{vvv}$. The weights estimated for these triads will indicate whether our prior expectations are validated. We also consider a single pairwise feature template,
a metric from \newcite{adamic2003friends} that sums over the mutual friends of $i$ and $j$, assigning more weight to mutual friends who themselves have a small number of friends:
\begin{equation}
AA(i,j) = \sum_{k \in \Gamma(i) \cap k \in \Gamma(j)} \frac{1}{\log \#|\Gamma(k)|},
\end{equation}
where $\Gamma(i)$ is the set of friends of node $i$. (We also tried simply counting the number of mutual friends, but the Adamic-Adar metric performs slightly better.) This feature appears in the vector $\vf(y_{ij},i,j,G)$, as defined in Equation~\ref{eq:prior}.

\section{Application to movie dialogues}
\label{sec:movies}
We apply the ideas in this paper to a dataset of movie dialogues~\cite{danescu2011chameleons}, including roughly 300,000 conversational turns between 10,000 pairs of characters in 617 movies. This dataset is chosen because it not only provides the script of each movie, but also indicates which characters are in dialogue in each line. We evaluate on quantitative measures of predictive likelihood (a token-level evaluation) and coherence of the induced address term clusters (a type-level evaluation). In addition, we describe in detail the inferred signed social networks on two films.

We evaluate the effects of three groups of features: address terms, mutual friends (using the Adamic-Adar metric), and triads. We include address terms in all evaluations, and test whether the network features improve performance. Ablating both network features is equivalent to clustering dyads by the counts of address terms, but all evaluations were performed by ablating components of the full model. We also tried ablating the text features, clustering edges using only the mutual friends and triad features, but we found that the resulting clusters were incoherent, with no discernible relationship to the address terms.

\subsection{Predictive log-likelihood}
To compute the predictive log-likelihood of the address terms, we hold out a randomly-selected 10\% of films. On these films, we use the first 50\% of address terms to estimate the dyad-label beliefs $q_{ij}(y)$. We then evaluate the expected log-likelihood of the second 50\% of address terms, computed as $\sum_y q_{ij}(y) \sum_n \log P(x_n \mid \theta_y)$ for each dyad. This is comparable to standard techniques for computing the held-out log-likelihood of topic models~\cite{wallach2009evaluation}.

As shown in Table~\ref{tab:lls}, the full model substantially outperforms the ablated alternatives. This indicates that the signed triad features contribute meaningful information towards the understanding of address terms in dialogue.

\begin{table}
  \centering
  \begin{tabular}{llll}
    \toprule
    \parbox{1.5cm}{Address\\terms} & \parbox{1.5cm}{Mutual\\friends} &     \parbox{1.5cm}{Signed\\triads} & Log-likelihood \\
    \midrule
    \checkmark &   &   & -2133.28 \\
    \checkmark &   & \checkmark & -2018.21 \\
    \checkmark & \checkmark &   & -1884.02 \\
    \checkmark & \checkmark & \checkmark & -1582.43 \\
    \bottomrule
  \end{tabular}
  \caption{Predictive log-likelihoods.}
  \label{tab:lls}
\end{table}

\subsection{Cluster coherence}
Next, we consider the model inferences that result when applying the EM procedure to the entire dataset. Table~\ref{tab:cluster-terms} presents the top address terms for each cluster, according to likelihood ratio. The cluster shown on the left emphasizes full names, titles, and formal address, while the cluster on the right includes the given name and informal address terms such as \example{man}, \example{baby}, and \example{dude}. We therefore use the labels ``\vos-cluster'' and ``\tu-cluster'', referring to the formal and informal clusters, respectively.

\begin{table}
  \centering
  \begin{tabular}{ll}
    \toprule
    \vos-cluster & \tu-cluster \\
    \midrule
\example{sir} & \annot{firstName}\\
\example{mr}+\annot{lastName} & \example{man}\\
\example{mr}+\annot{firstName} & \example{baby}\\
\example{mr} & \example{honey}\\
\example{miss}+\annot{lastName} & \example{darling}\\
\example{son} & \example{sweetheart}\\
\example{mister}+\annot{firstName} & \example{buddy}\\
\example{mrs} & \example{sweetie}\\
\example{mrs}+\annot{lastName} & \example{hon}\\
\annot{firstName}+\annot{lastName} & \example{dude}\\
    \bottomrule
  \end{tabular}
  \caption{The ten strongest address terms for each cluster, sorted by likelihood ratio.}
\label{tab:cluster-terms}
\end{table}

We perform a quantitative evaluation of this clustering through an intrusion task~\cite{chang2009reading}. Specifically, we show individual raters three terms, selected so that two terms are from the same cluster, and the third term is from the other cluster; we then ask them to identify which term is least like the other two. Five raters were each given a list of forty triples, with the order randomized. Of the forty triples, twenty were from our full model, and twenty were from a text-only clustering model. The raters agreed with our full model in 73\% percent of cases, and agreed with the text-only model in 52\% percent of cases. By Fisher's exact test, this difference is statistically significant at $p < 0.01$. Both results are significantly greater than chance agreement (33\%) by a binomial test, $p < 0.001$.

\subsection{Network feature weights}
\begin{figure}
  \centering
  \includegraphics[width=0.5\textwidth]{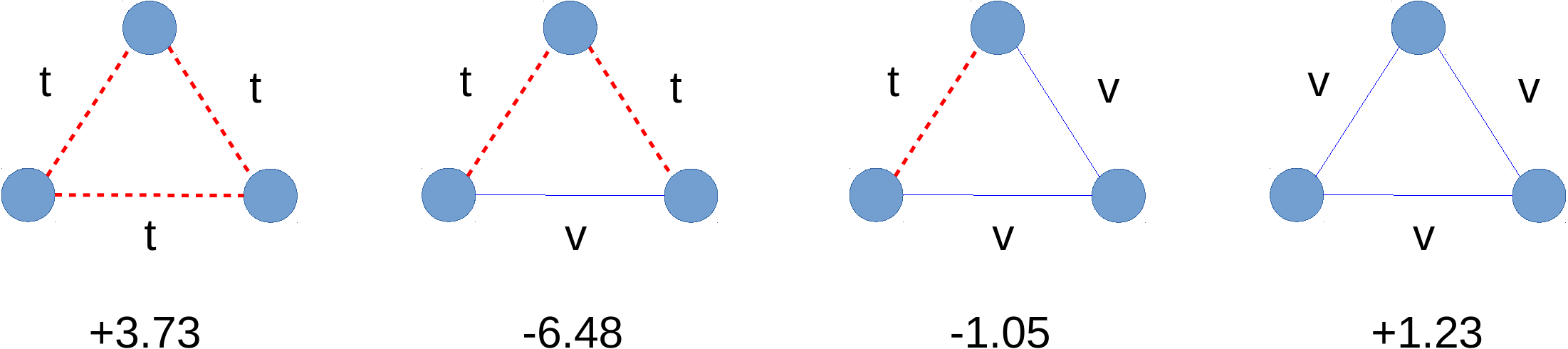}
  \caption{Estimated triad feature weights}
  \label{fig:triads}
\end{figure}

Figure~\ref{fig:triads} shows the feature weights for each of the four possible triads. Triads with homogeneous signs are preferred, particularly \annot{ttt} (all informal); heterogeneous triads are dispreferred, particularly \annot{ttv}, which is when two individuals have a formal relationship despite having a mutual informal tie. Less dispreferred is \annot{tvv}, when a pair of friends have an informal relationship despite both having a formal relationship with a third person; consider, for example, the situation of two students and their professor. In addition, the informal sign is preferred when the dyad has a high score on the Adamic-Adar metric, and dispreferred otherwise. This coheres with the intuition that highly-embedded edges are likely to be informal, with many shared friends.

\subsection{Qualitative results}
\begin{figure}
  \centering
  \includegraphics[width=0.5\textwidth]{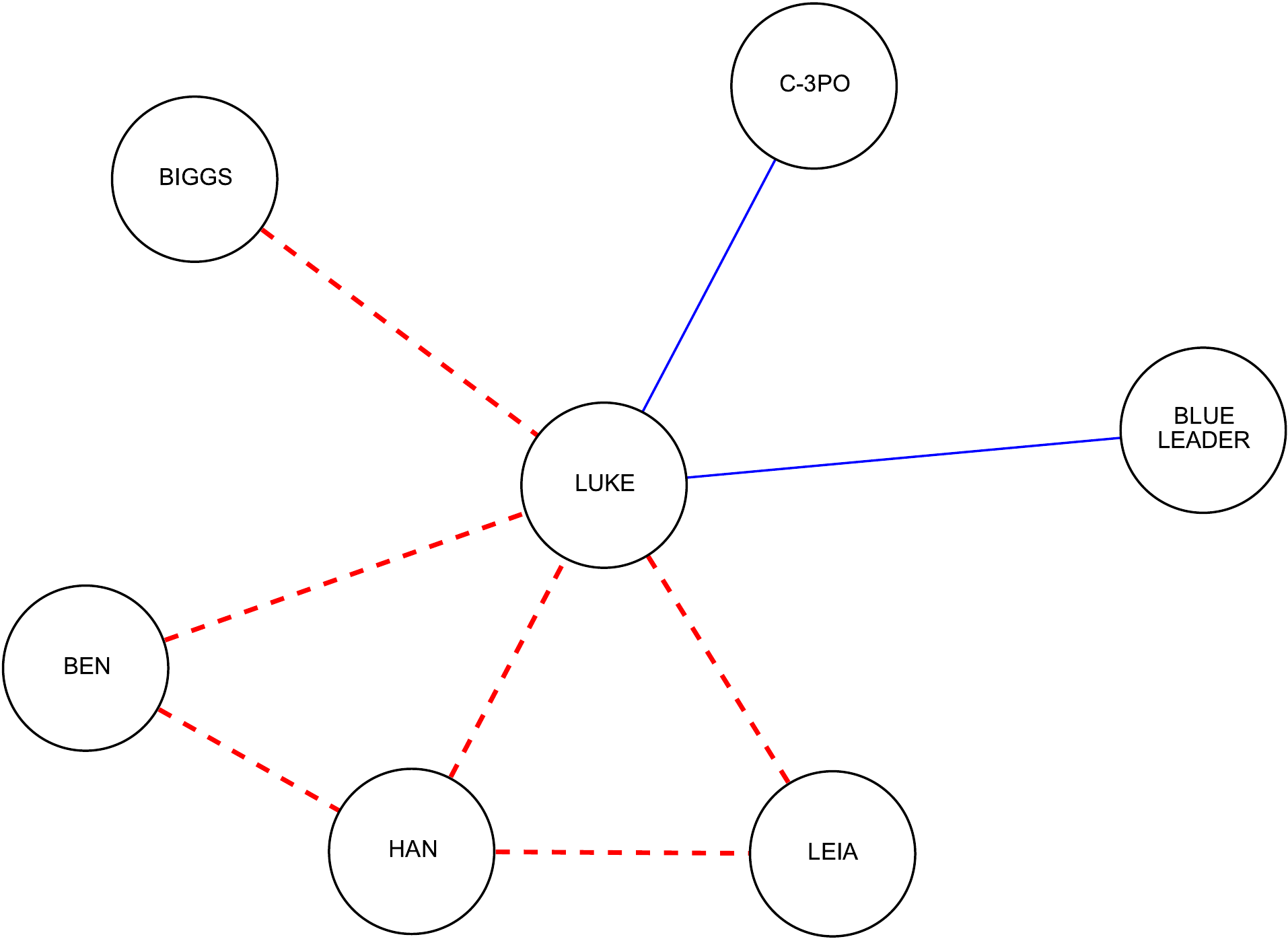}
  \caption{Induced signed social network from the film \emph{Star Wars}. Blue solid edges are in the \vos-cluster, red dashed edges are in the \tu-cluster.}
  \label{fig:starwars}
\end{figure}

Analysis of individual movies suggests that the induced tie signs are meaningful and coherent. For example, the film ``Star Wars'' is a space opera, in which the protagonists Luke, Han, and Leia attempt to defeat an evil empire led by Darth Vader. The induced signed social network is shown in Figure~\ref{fig:starwars}. The \vos-edges seem reasonable: C-3PO is a robotic servant, and Blue Leader is Luke's military commander (\textsc{Blue leader}: \example{Forget it, son.} \textsc{Luke}: \example{Yes, sir, but I can get him...}). In contrast, the character pairs with \tu-edges all have informal relationships: the lesser-known character Biggs is Luke's more experienced friend (\textsc{Biggs}: \example{That's no battle, kid}).

\begin{figure}
  \centering
  \includegraphics[width=0.5\textwidth]{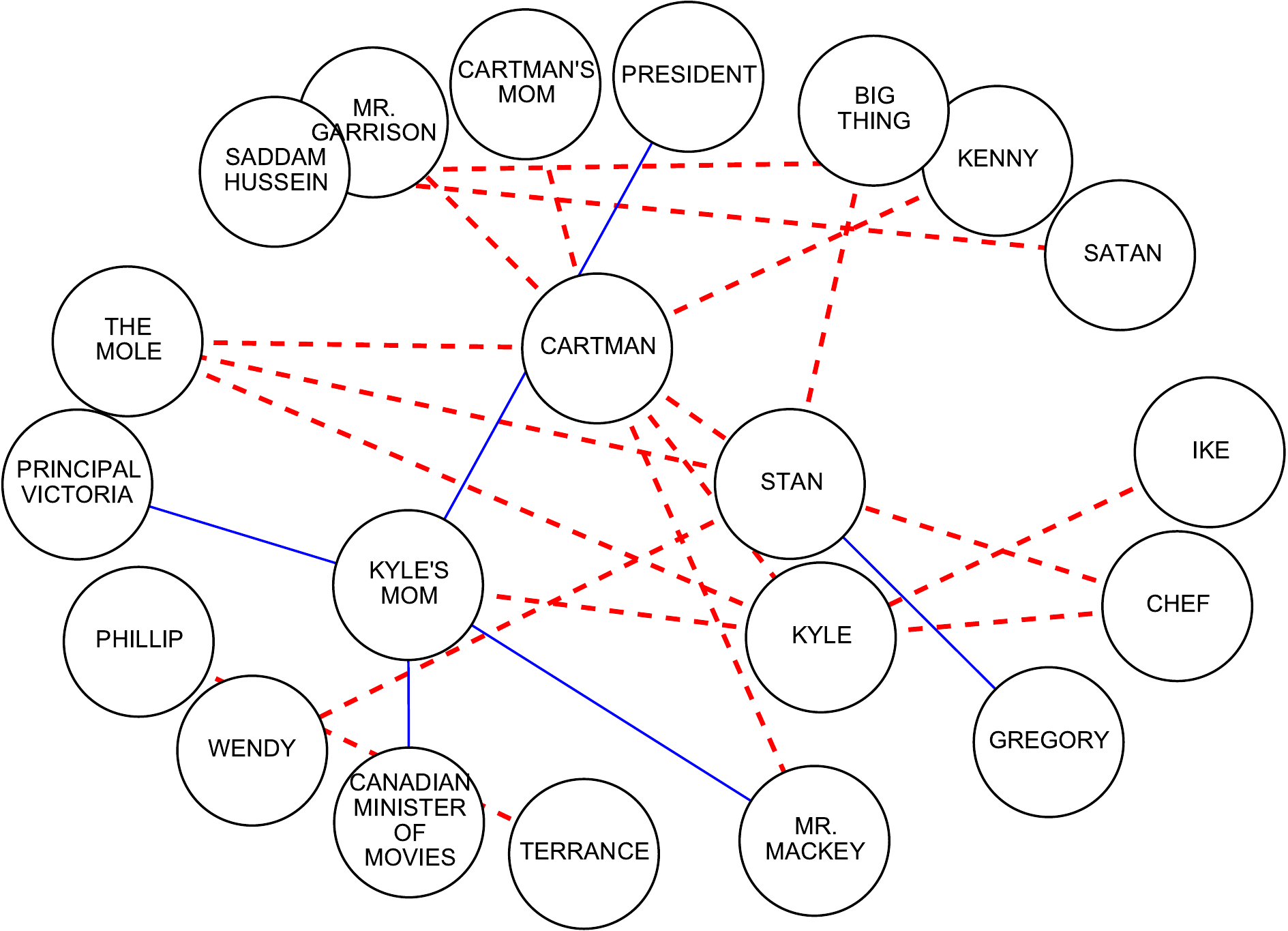}
  \caption{Induced signed social network from the film \emph{South Park: Bigger, Longer \& Uncut}. Blue solid edges are in the \vos-cluster, red dashed edges are in the \tu-cluster.}
  \label{fig:south-park}
\end{figure}

The animated film ``South Park: Bigger, Longer \& Uncut'' centers on three children: Stan, Cartman, and Kyle; it also involves their parents, teachers, and friends, as well as a number of political and religious figures. The induced social network is shown in Figure~\ref{fig:south-park}. The children and their associates mostly have \tu-edges, except for the edge to Gregory, a British character with few speaking turns. This part of the network also has a higher clustering coefficient, as the main characters share friends such as Chef and The Mole. The left side of the diagram centers on Kyle's mother, who has more formal relationships with a variety of authority figures.

\section{Related work}
Recent work has explored the application of signed social network models to social media. \newcite{leskovec2010signed} find three social media datasets from which they are able to identify edge polarity; this enables them to compare the frequency of signed triads against baseline expectations, and to build a classifier to predict edge labels~\cite{leskovec2010predicting}. However, in many of the most popular social media platforms, such as Twitter and Facebook, there is no metadata describing edge labels. We are also interested in new applications of signed social network analysis to datasets outside the realm of social media, such as literary texts~\cite{moretti2005graphs,elson2010extracting,agarwal2013automatic} and movie scripts, but in such corpora, edge labels are not easily available.

In many datasets, it is possible to obtain the textual content exchanged between members of the network, and this content can provide a signal for network structure. For example, \newcite{hassan2012extracting} characterize the sign of each network edge in terms of the \textbf{sentiment} expressed across it, finding that the resulting networks cohere with the predictions of structural balance theory; similar results are obtained by~\newcite{west2014exploiting}, who are thereby able to predict the signs of unlabeled ties. Both papers leverage the relatively mature technology of sentiment analysis, and are restricted to edge labels that reflect sentiment. The unsupervised approach presented here could in principle be applied to lexicons of sentiment terms, rather than address terms, but we leave this for future work.

The issue of address formality in English was considered by~\newcite{faruqui2011thou}, who show that annotators can label the formality of the second person pronoun with agreement of 70\%. They use these annotations to train a supervised classifier, obtaining comparable accuracy. If no labeled data is available, annotations can be projected from languages where the T/V distinction is marked in the morphology of the second person pronoun, such as German~\cite{faruqui2012towards}. Our work shows that it is possible to detect formality without labeled data or parallel text, by leveraging regularities across network structures; however, this requires the assumption that the level of formality for a pair of individuals is constant over time. The combination of our unsupervised approach with annotation projection might yield models that attain higher performance while capturing change in formality over time.

More broadly, a number of recent papers have proposed to detect various types of social relationships from linguistic content. Of particular interest are power relationships, which can be induced from n-gram features~\cite{bramsen2011extracting,prabhakaran2012predicting} and from \jemph{coordination}, where one participant's linguistic style is asymmetrically affected by the other~\cite{danescu2012echoes}. \newcite{danescu2013computational} describe an approach to recognizing politeness in text, lexical and syntactic features motivated by politeness theory. \newcite{anand2011cats} detect ``rebuttals'' in argumentative dialogues, and \newcite{hasan2013extra} employ extra-linguistic structural features to improve the detection of stances in such debates. In all of these cases, labeled data is used to train supervised model; our work shows that social structural regularities are powerful enough to support accurate induction of social relationships (and their linguistic correlates) without labeled data.
\section{Conclusion}
This paper represents a step towards unifying theoretical models of signed social network structures with linguistic accounts of the expression of social relationships in dialogue. By fusing these two phenomena into a joint probabilistic model, we can induce edge types with robust linguistic signatures and coherent structural properties.  
We demonstrate the effectiveness of this approach on movie dialogues, where it induces symmetric \tu/\vos~networks and their linguistic signatures without supervision. Future work should evaluate the capability of this approach to induce asymmetric signed networks, the utility of partial or distant supervision, and applications to non-fictional dialogues.

\section*{Acknowledgments}
We thank the reviewers for their detailed feedback. The paper benefitted from conversations with Cristian Danescu-Niculescu-Mizil, Chris Dyer, Johan Ugander, and Bob West. This research was supported by an award from the Air Force Office of Scientific Research, and by Google, through a Focused Research Award for Computational Journalism.

\bibliographystyle{naaclhlt2015}

\begin{thebibliography}{}

\bibitem[\protect\citename{Adamic and Adar}2003]{adamic2003friends}
Lada~A Adamic and Eytan Adar.
\newblock 2003.
\newblock Friends and neighbors on the web.
\newblock {\em Social networks}, 25(3):211--230.

\bibitem[\protect\citename{Agarwal \bgroup et al.\egroup
  }2013]{agarwal2013automatic}
Apoorv Agarwal, Anup Kotalwar, and Owen Rambow.
\newblock 2013.
\newblock Automatic extraction of social networks from literary text: A case
  study on alice in wonderland.
\newblock In {\em the Proceedings of the 6th International Joint Conference on
  Natural Language Processing (IJCNLP 2013)}.

\bibitem[\protect\citename{Anand \bgroup et al.\egroup }2011]{anand2011cats}
Pranav Anand, Marilyn Walker, Rob Abbott, Jean~E. Fox~Tree, Robeson Bowmani,
  and Michael Minor.
\newblock 2011.
\newblock Cats rule and dogs drool!: Classifying stance in online debate.
\newblock In {\em Proceedings of the 2nd Workshop on Computational Approaches
  to Subjectivity and Sentiment Analysis (WASSA 2.011)}, pages 1--9, Portland,
  Oregon, June. Association for Computational Linguistics.

\bibitem[\protect\citename{Bramsen \bgroup et al.\egroup
  }2011]{bramsen2011extracting}
Philip Bramsen, Martha Escobar-Molano, Ami Patel, and Rafael Alonso.
\newblock 2011.
\newblock Extracting social power relationships from natural language.
\newblock In {\em {Proceedings of the Association for Computational Linguistics
  (ACL)}}, pages 773--782, Portland, OR.

\bibitem[\protect\citename{Brown and Ford}1961]{brown1961address}
Roger Brown and Marguerite Ford.
\newblock 1961.
\newblock Address in american english.
\newblock {\em The Journal of Abnormal and Social Psychology}, 62(2):375.

\bibitem[\protect\citename{Brown}1987]{brown1987politeness}
Penelope Brown.
\newblock 1987.
\newblock {\em Politeness: Some universals in language usage}, volume~4.
\newblock Cambridge University Press.

\bibitem[\protect\citename{Cartwright and
  Harary}1956]{cartwright1956structural}
Dorwin Cartwright and Frank Harary.
\newblock 1956.
\newblock Structural balance: a generalization of heider's theory.
\newblock {\em Psychological review}, 63(5):277.

\bibitem[\protect\citename{Chang \bgroup et al.\egroup }2009]{chang2009reading}
Jonathan Chang, Sean Gerrish, Chong Wang, Jordan~L Boyd-graber, and David~M
  Blei.
\newblock 2009.
\newblock Reading tea leaves: How humans interpret topic models.
\newblock In {\em {Neural Information Processing Systems (NIPS)}}, pages
  288--296.

\bibitem[\protect\citename{Danescu-Niculescu-Mizil and
  Lee}2011]{danescu2011chameleons}
Cristian Danescu-Niculescu-Mizil and Lillian Lee.
\newblock 2011.
\newblock Chameleons in imagined conversations: A new approach to understanding
  coordination of linguistic style in dialogs.
\newblock In {\em Proceedings of the {ACL} Workshop on Cognitive Modeling and
  Computational Linguistics}.

\bibitem[\protect\citename{Danescu-Niculescu-Mizil \bgroup et al.\egroup
  }2012]{danescu2012echoes}
Cristian Danescu-Niculescu-Mizil, Lillian Lee, Bo~Pang, and Jon Kleinberg.
\newblock 2012.
\newblock Echoes of power: Language effects and power differences in social
  interaction.
\newblock In {\em {Proceedings of the Conference on World-Wide Web (WWW)}},
  pages 699--708, Lyon, France.

\bibitem[\protect\citename{Danescu-Niculescu-Mizil \bgroup et al.\egroup
  }2013]{danescu2013computational}
Cristian Danescu-Niculescu-Mizil, Moritz Sudhof, Dan Jurafsky, Jure Leskovec,
  and Christopher Potts.
\newblock 2013.
\newblock A computational approach to politeness with application to social
  factors.
\newblock In {\em {Proceedings of the Association for Computational Linguistics
  (ACL)}}, pages 250--259, Sophia, Bulgaria.

\bibitem[\protect\citename{Dempster \bgroup et al.\egroup
  }1977]{dempster1977maximum}
Arthur~P Dempster, Nan~M Laird, and Donald~B Rubin.
\newblock 1977.
\newblock Maximum likelihood from incomplete data via the em algorithm.
\newblock {\em Journal of the Royal Statistical Society. Series B
  (Methodological)}, pages 1--38.

\bibitem[\protect\citename{Elson \bgroup et al.\egroup
  }2010]{elson2010extracting}
David~K Elson, Nicholas Dames, and Kathleen~R McKeown.
\newblock 2010.
\newblock Extracting social networks from literary fiction.
\newblock In {\em {Proceedings of the Association for Computational Linguistics
  (ACL)}}, pages 138--147, Uppsala, Sweden.

\bibitem[\protect\citename{Faruqui and Pad\'{o}}2011]{faruqui2011thou}
Manaal Faruqui and Sebastian Pad\'{o}.
\newblock 2011.
\newblock {"I Thou Thee, Thou Traitor": Predicting Formal vs. Informal Address
  in English Literature}.
\newblock In {\em {Proceedings of the Association for Computational Linguistics
  (ACL)}}, pages 467--472, Portland, OR.

\bibitem[\protect\citename{Faruqui and Pad\'{o}}2012]{faruqui2012towards}
Manaal Faruqui and Sebastian Pad\'{o}.
\newblock 2012.
\newblock Towards a model of formal and informal address in english.
\newblock In {\em {Proceedings of the European Chapter of the Association for
  Computational Linguistics (EACL)}}, pages 623--633.

\bibitem[\protect\citename{Gutmann and Hyv{\"a}rinen}2012]{gutmann2012noise}
Michael~U Gutmann and Aapo Hyv{\"a}rinen.
\newblock 2012.
\newblock Noise-contrastive estimation of unnormalized statistical models, with
  applications to natural image statistics.
\newblock {\em The Journal of Machine Learning Research}, 13(1):307--361.

\bibitem[\protect\citename{Hasan and Ng}2013]{hasan2013extra}
Kazi~Saidul Hasan and Vincent Ng.
\newblock 2013.
\newblock Extra-linguistic constraints on stance recognition in ideological
  debates.
\newblock In {\em {Proceedings of the Association for Computational Linguistics
  (ACL)}}, pages 816--821, Sophia, Bulgaria.

\bibitem[\protect\citename{Hassan \bgroup et al.\egroup
  }2012]{hassan2012extracting}
Ahmed Hassan, Amjad Abu-Jbara, and Dragomir Radev.
\newblock 2012.
\newblock Extracting signed social networks from text.
\newblock In {\em Workshop Proceedings of TextGraphs-7 on Graph-based Methods
  for Natural Language Processing}, pages 6--14. Association for Computational
  Linguistics.

\bibitem[\protect\citename{Hymes}1972]{hymes1972communicative}
Dell Hymes.
\newblock 1972.
\newblock On communicative competence.
\newblock {\em Sociolinguistics}, pages 269--293.

\bibitem[\protect\citename{Kiesling}2004]{kiesling2004dude}
Scott~F Kiesling.
\newblock 2004.
\newblock Dude.
\newblock {\em American Speech}, 79(3):281--305.

\bibitem[\protect\citename{Leskovec \bgroup et al.\egroup
  }2010a]{leskovec2010predicting}
Jure Leskovec, Daniel Huttenlocher, and Jon Kleinberg.
\newblock 2010a.
\newblock Predicting positive and negative links in online social networks.
\newblock In {\em {Proceedings of the Conference on World-Wide Web (WWW)}},
  pages 641--650.

\bibitem[\protect\citename{Leskovec \bgroup et al.\egroup
  }2010b]{leskovec2010signed}
Jure Leskovec, Daniel Huttenlocher, and Jon Kleinberg.
\newblock 2010b.
\newblock Signed networks in social media.
\newblock In {\em {Proceedings of Human Factors in Computing Systems (CHI)}},
  pages 1361--1370.

\bibitem[\protect\citename{Liben-Nowell and Kleinberg}2007]{liben2007link}
David Liben-Nowell and Jon Kleinberg.
\newblock 2007.
\newblock The link-prediction problem for social networks.
\newblock {\em Journal of the American society for information science and
  technology}, 58(7):1019--1031.

\bibitem[\protect\citename{Liu and Nocedal}1989]{liu1989limited}
Dong~C Liu and Jorge Nocedal.
\newblock 1989.
\newblock {On the limited memory BFGS method for large scale optimization}.
\newblock {\em Mathematical programming}, 45(1-3):503--528.

\bibitem[\protect\citename{Manning \bgroup et al.\egroup
  }2014]{manning2014stanford}
Christopher~D. Manning, Mihai Surdeanu, John Bauer, Jenny Finkel, Steven~J.
  Bethard, and David McClosky.
\newblock 2014.
\newblock The {Stanford} {CoreNLP} natural language processing toolkit.
\newblock In {\em Proceedings of 52nd Annual Meeting of the Association for
  Computational Linguistics: System Demonstrations}, pages 55--60.

\bibitem[\protect\citename{Marcus \bgroup et al.\egroup
  }1993]{marcus1993building}
Mitchell~P Marcus, Mary~Ann Marcinkiewicz, and Beatrice Santorini.
\newblock 1993.
\newblock Building a large annotated corpus of {English}: {The Penn Treebank}.
\newblock {\em Computational Linguistics}, 19(2):313--330.

\bibitem[\protect\citename{Moretti}2005]{moretti2005graphs}
Franco Moretti.
\newblock 2005.
\newblock {\em Graphs, maps, trees: abstract models for a literary history}.
\newblock Verso.

\bibitem[\protect\citename{Prabhakaran \bgroup et al.\egroup
  }2012]{prabhakaran2012predicting}
Vinodkumar Prabhakaran, Owen Rambow, and Mona Diab.
\newblock 2012.
\newblock Predicting overt display of power in written dialogs.
\newblock In {\em {Proceedings of the North American Chapter of the Association
  for Computational Linguistics (NAACL)}}, pages 518--522.

\bibitem[\protect\citename{Ratinov and Roth}2009]{ratinov2009design}
Lev Ratinov and Dan Roth.
\newblock 2009.
\newblock Design challenges and misconceptions in named entity recognition.
\newblock In {\em Proceedings of the Thirteenth Conference on Computational
  Natural Language Learning}, pages 147--155. Association for Computational
  Linguistics.

\bibitem[\protect\citename{Wainwright and Jordan}2008]{wainwright2008graphical}
Martin~J Wainwright and Michael~I Jordan.
\newblock 2008.
\newblock Graphical models, exponential families, and variational inference.
\newblock {\em Foundations and Trends{\textregistered} in Machine Learning},
  1(1-2):1--305.

\bibitem[\protect\citename{Wallach \bgroup et al.\egroup
  }2009]{wallach2009evaluation}
Hanna~M Wallach, Iain Murray, Ruslan Salakhutdinov, and David Mimno.
\newblock 2009.
\newblock Evaluation methods for topic models.
\newblock In {\em {Proceedings of the International Conference on Machine
  Learning (ICML)}}, pages 1105--1112.

\bibitem[\protect\citename{West \bgroup et al.\egroup
  }2014]{west2014exploiting}
Robert West, Hristo Paskov, Jure Leskovec, and Christopher Potts.
\newblock 2014.
\newblock Exploiting social network structure for person-to-person sentiment
  analysis.
\newblock {\em Transactions of the Association for Computational Linguistics},
  2:297--310.

\end{thebibliography}
\citestr

\end{document}